\documentclass[conference]{IEEEtran}
\usepackage{graphicx,epsfig,color,graphics}
\newtheorem{theorem}{Theorem}
\newtheorem{lemma}{Lemma}

\begin{document}

\title{A Combinatorial Family of Near Regular LDPC Codes}




\author{\authorblockN{K. Murali Krishnan, Rajdeep Singh, 
L. Sunil Chandran and Priti Shankar }
  \authorblockA{Department of Computer Science and Automation\\
                Indian Institute of Science, Bangalore 560012, \\ India.}}


%


\maketitle

\begin{abstract}
An elementary combinatorial Tanner graph construction for a family of
near-regular low density parity check (LDPC) codes achieving high girth 
is presented.  The construction allows flexibility in the choice of 
design parameters like rate, average degree,  girth and block length 
of the code and yields an asymptotic family.  
The complexity of constructing codes in the family
grows only quadratically with the block length.  
\end{abstract}

\IEEEpeerreviewmaketitle

\section{Introduction}
\label{intro}

The fact that iterative decoding on LDPC codes performs well when the
underlying Tanner graph \cite{Tan} has large girth is well known \cite{Gal}.  
The recent revival of interest
in LDPC codes owing to their near capacity performance on various channel
models has resulted in considerable research on the construction of 
LDPC code families of high rate and large girth.   These
constructions may be classified as 
random code constructions (for example see \cite{Di, Luby}), 
construction of codes based on projective and combinatorial 
geometries (see \cite{Lin, Vas, Tang}
and references therein),  heuristic search based constructions
\cite{Tian, Adi}, constructions based on 
circulant matrices \cite{Sri2}, 
algebraic constructions (see \cite{ Ros, Sri1}),  
code constructions based on expander graphs \cite{Sip,Zem}, and edge growth constructions ~\cite{Hu}.  

In this note, we present an elementary graph theoretic construction
for a family of binary LDPC codes. 
These codes achieve high girth and are 
almost regular in the sense that the degree of a vertex is allowed 
to differ by at most one from the average.  We shall refer to these
codes as {\em ARG} (Almost Regular high Girth) codes.  
The construction gives flexibility in the 
design parameters of the code 
like rate, block-length, and average degree, and yields an asymptotic family.
We prove bounds on code parameters achieved by the construction. 
The complexity of the graph construction algorithm
grows only quadratically with the block length of the code.  

The construction here is similar in spirit to the very general graph 
construction scheme called the progressive edge-growth (PEG) algorithm
proposed in \cite{Hu} and may be considered as being
specially tuned for obtaining near regular graphs of large girth.  
However in \cite{Hu} no technique for simultaneously bounding 
the maximum left and right degrees of the graphs constructed is provided, 
and hence the girth bounds depend on the values of the degrees obtained 
experimentally.  The authors report that good values of girth can 
be achieved in practice. 

The bounds on the node degrees in the 
Tanner graph construction proposed here are achieved
by adapting a high girth graph construction technique known in 
the graph theory literature \cite{Sunil}.  The following section
presents the construction and establishes the bound on the girth
of the Tanner graph constructed.  Simulations indicate that  
rate $1/2$ ARG codes perform better than regular
codes of the same block length reported in \cite{Mc}.

\section{The Code Construction}
\label{reduction}

Given a bipartite graph $G=(L,R,E\subseteq L\times R)$, 
$|L|=n$, $|R|=m$, the $m\times n$ parity check matrix $H(G)=[h_{i,j}]$ defined 
by $h_{i,j}=1$ if and only if $(j,i)\in E$, $1\leq j\leq n$, $1\leq i\leq m$
specifies a binary linear code $C(G)$.  
We say $G$ is the {\em Tanner graph} for $C(G)$. The code $C(G)$ is an LDPC 
code if the maximum degree of any vertex in $G$ is bounded by a constant.
The length of the shortest cycle in $G$ is called the {\em girth} of $G$
denoted by $g(G)$.  In the following, we describe the construction of a 
bipartite Tanner graph and give bounds on the parameters of the code defined 
by the graph.

Let $m,n,p,q$ be positive integers with $n>m>1$, $p<q$, $np=mq$ and
let $d$ be constant with $d\leq (m+3)/3(p+q)$. 
We construct a bipartite graph $G=(L,R,E)$ 
with average left degree $dp$ and average right degree $dq$ as follows.  
Initially $L=\{1..n\}$, $R=\{1..m\}$ and $E=\emptyset$.  
We denote by $deg(x)$ the degree of a vertex $x\in L\cup R$. 
Denote by $\delta (x,y)$ the length of the shortest path from $x$ to $y$
in $G$.  Clearly $deg(x)=0$  and $\delta (x,y)=\infty$
for all $x,y\in L\cup R$ initially.

We will add $npd(=mqd)$ edges to $G$ one by one.  When the $e^{th}$ edge
is added for some $1\leq e\leq npd$ we shall say that the algorithm is 
in phase $(i,j)$ where $i=\lceil e/n\rceil$ and $j=\lceil e/m\rceil$. 
We say that the edge belongs to {\em left phase} $i$ and 
{\em right phase} $j$.  Thus the first $m$ edges will be added 
during phase $(1,1)$, edges $m+1$ to $\min \{n,2m\}$  
will be added during phase $(1,2)$ and so on.  Note that after
left phase $i$, the average left degree of the graph will be $i$.
Similarly, the average right degree will be $j$ at the end of 
right phase $j$.  The algorithm terminates at the end of phase $(dp,dq)$.

The algorithm repeatedly picks up a vertex of minimum degree (chosen
alternately from $L$ and $R$) and adds from it an edge 
to the farthest vertex on the opposite side in such a way that the 
vertex degrees are not allowed to become excessive. 
During phase $(i,j)$, the degree of a left vertex never exceeds $i+1$
and the degree of a right vertex never exceeds $j+1$.
We will prove that at the end of left phase $i$,
every vertex in $L$ has degree at least $i-1$ and at the end of right
phase $j$ every vertex in $R$ has degree at least $j-1$.
Hence, when the algorithm terminates, the left and
the right degrees are bounded above by $pd+1$ and $qd+1$ respectively,
and bounded below by $pd-1$ and $qd-1$ respectively yielding a 
near-regular graph.  The steps of the algorithm are formalized below: \\

\begin{itemize}
\item for $e$ $:=$ $1$ to $npd$ do  \hspace{0.3 in} \{$npd=mqd$ edges added \}

\item \begin{enumerate}

       \item $i:=\lceil e/n\rceil$ \hspace{0.1 in} 
             $j:=\lceil e/m\rceil$ \hspace{0.5 in} \{phase is $(i,j)$\}

       \item if $e$ is {\em odd}, choose a vertex $x$ of minimum degree from 
             $L$. Let $S=\{z\in R:\delta (x,z)>1$ and $deg(z)<j+1\}$. 
             Select a $y\in S$ such that $\delta (x,y)\geq \delta(x,z)$ 
             for all $z\in S$.  Add $(x,y)$ to $E$.  
               
       \item else if $e$ is {\em even}, choose a vertex $x$ of minimum degree 
             from $R$. Let $S=\{z\in L:\delta (x,z)>1$ and $deg(z)<i+1\}$. 
             Select a $y\in S$ such that $\delta (x,y)\geq \delta(x,z)$. 
             for all $z\in S$. Add $(x,y)$ to $E$.  \\
       \end{enumerate}
\end{itemize}

We shall call edges corresponding to odd and even values of $e$ as 
{\em odd edges} and {\em even edges} respectively.
Note that the algorithm may fail to progress if the set $S$ becomes empty
and no edge could be added during some intermediate phase.  We shall
rule out this possibility later.

\begin{theorem} 
$C(G)$ is an LDPC code with rate $r\geq 1-p/q$.
\end{theorem}
\begin{proof}
Since $H(G)$ is an $m\times n$ matrix, $r\geq 1-m/n$.  Since $m/n=p/q$ 
by assumption, the claim on rate follows. By construction the left and
right degrees of any node in $G$ are bounded by $pd+1$ and $qd+1$. Since
$d$ is constant the graph is of low density. 
\end{proof}

The following lemma proved by induction establishes the key invariants 
maintained by the algorithm.

\begin{lemma}
For all $1\leq i\leq pd$ and $1\leq j\leq qd$ the following holds:
\begin{itemize}
\item If the algorithm completes left phase $i$  
then $i-1\leq deg(x)\leq i+1$ for all $x\in L$ 
at the end of left phase $i$. 
\item If the algorithm completes right phase $j$
then $j-1\leq deg(y)\leq j+1$ for all $y\in R$
at the end of right phase $j$. 
\end{itemize}
\end{lemma}

\begin{proof}
We shall prove the first statement using induction.  
Initially the hypothesis holds.  
Assume the statement true for some $i$, $0\leq i< pd$ and consider the 
the situation after completion of left phase $i+1$. 
Let $n^-$, $n^0$ and $n^+$ be the number of vertices with degree 
$i-1$, $i$ and $i+1$ respectively at the end of left phase $i$.  
Since the average degree of a left node is $i$ at 
the end of left phase $i$, we have the following:
\begin{eqnarray}
(i-1)n^{-}+in^{0}+(i+1)n^{+}=in=i(n^{-}+n^{0}+n^{+}).
\end{eqnarray}
Canceling terms we have $n^{-}=n^{+}\leq \lfloor n/2 \rfloor$.  
Thus to satisfy the lower bound in the induction hypothesis 
at most $\lfloor n/2\rfloor$
edges need to be added to the $n^{-}$ deficient vertices in $L$ during 
left phase $i+1$.  Since out of the $n$ edges added during left phase 
$i+1$ at least $\lfloor n/2\rfloor$ must be from minimum degree vertices in 
$L$ (because every odd edge will be added from a
vertex of minimum degree in $L$), all the $n^{-}$ deficient vertices
would have increased their degree by at least one and the lower bound on 
the left degree will be satisfied after phase $i+1$.  Since the
average degree of a left vertex at the end of left phase $i+1$ is $i+1$, 
there always will exist a vertex $x$ in $L$ with $deg(x)<i+1$ before 
the completion of left phase $i+1$. Hence the algorithm
will never choose a left vertex of degree $i+1$ for adding an edge 
when an odd edge is added during phase $i+1$.  
Finally, the algorithm explicitly ensures that an even edge is
added from a vertex $y\in R$ to $x\in L$ during phase $i+1$ only if
$deg(x)\leq i+1$ before the addition.  Hence in all cases, the upper bound
on vertex degree is also maintained during left phase $i+1$. 
The second statement in the lemma is proved similarly.
\end{proof}

It remains to be shown that the algorithm will indeed complete
all the phases  successfully. The algorithm may fail to complete phase $i$
if at some stage the set $S$ constructed by the algorithm is empty.
The following lemma rules out this possibility.

\begin{lemma}
If $d\leq (m+3)/3(p+q)$ the algorithm will complete all the phases.   
\end{lemma}
\begin{proof}
Suppose that the algorithm fails at some phase $(i,j)$
because the set $S$ becomes empty while trying to add an odd edge from
a vertex $x\in L$.  By Lemma 1, $x$ must have at least $i-2$ neighbours,
each of degree at least $j-2$. Since $x$ has at most $i+1$ neighbours
(by Lemma 1) and $S=\emptyset$, there must be at least $m-i-1$ 
non-neigbours of $x$ in $R$ with degree $j+1$.
Thus the total degree of all vertices in $R$ must be
at least $(m-i-1)(j+1)+(i-2)(j-2)$.  However, before phase $(i,j)$ ends
the average right degree is less than $j$.  Hence we have:
\begin{eqnarray}
(m-i-1)(j+1)+(i-2)(j-2)<mj
\end{eqnarray}
After simplification this yields $(m+3)/3<(i+j)$.  Considering the case
when an even edge is added and applying similar arguments we get
the condition $(n+3)/3<(i+j)$.  Since $i\leq pd$, $j\leq qd$ and $m<n$,
if $d<(m+3)/3(p+q)$ the failure condition will never occur and the algorithm
will successfully complete phase $(pd,qd)$.
\end{proof}

We are now ready to prove the bound on the girth.

\begin{theorem}
$g(G)\geq 2\log_{pqd^2} (1+ m(pqd^2-1)/2(pd+1))$.
\end{theorem}
\begin{proof}
Assume that a smallest length cycle in $G$ of length $g(G)=2r$ was 
formed during phase $(i,j)$ of the algorithm. Assume   
$x\in L$ had the least degree and 
was connected to $y\in R$ during the addition of an odd edge
causing the cycle.  Let $T=\{z\in R:\delta (x,z)\geq g\}$.  
The node $x$ had to be connected to $y$ 
and not to any node in $T$ because $deg(z)=j+1$ for all $z\in T$. 
But there are at most $m/2$ nodes of degree $j+1$ during right 
phase $j$. Thus $|T|\leq m/2$. Hence $|R-T|\geq m/2$.  
But all nodes in $R-T$ must be at a distance 
at most $g-1=2r-1$ from $x$.
Since the maximum left and right degrees of a node in $G$ are bounded
by $pd+1$ and $qd+1$ respectively, the number of such nodes is bounded above by 
$(pd+1)+(pd+1)(pqd^2)+...+(pd+1)(pqd^2)^{r-1}
= (pd+1)((pqd^2)^{r}-1)/(pqd^2-1)$.
Combining the lower and upper bounds we get:

\begin{figure*}
  \begin{center}
	     \includegraphics[width=5.35in]{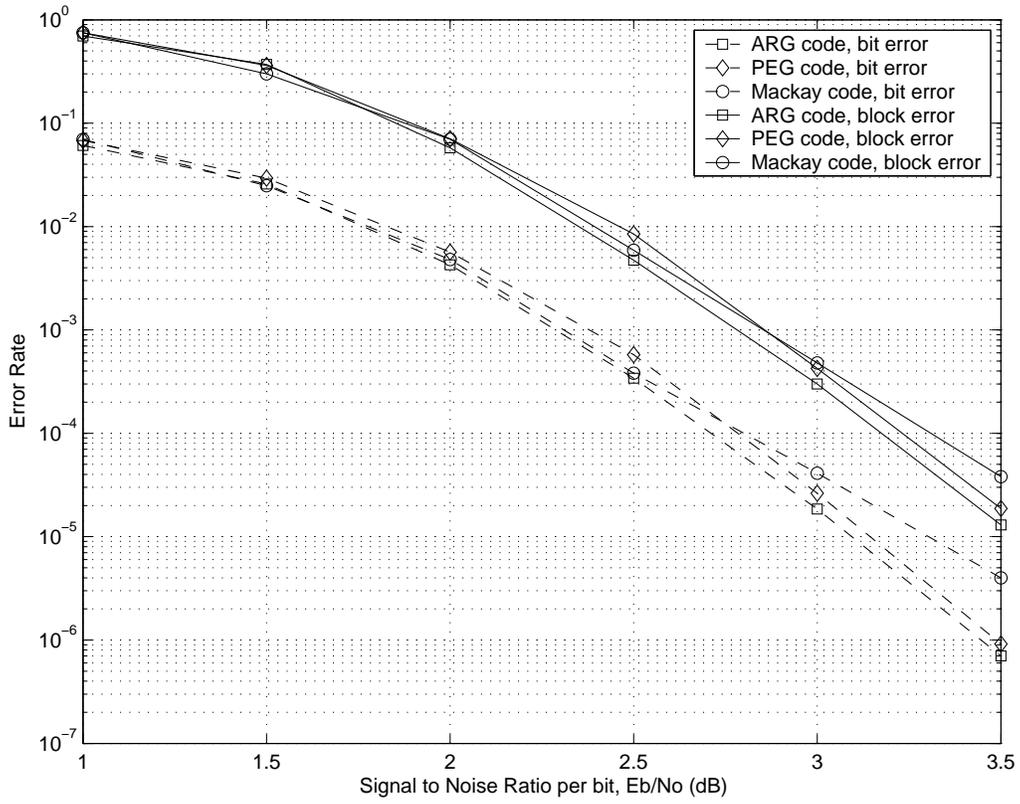}
	   \caption{Performance of ARG (504,8,3) code}
       \label{fig504}
   \end{center}
\end{figure*}

\begin{figure*}
  \begin{center}
	     \includegraphics[width=5.35in]{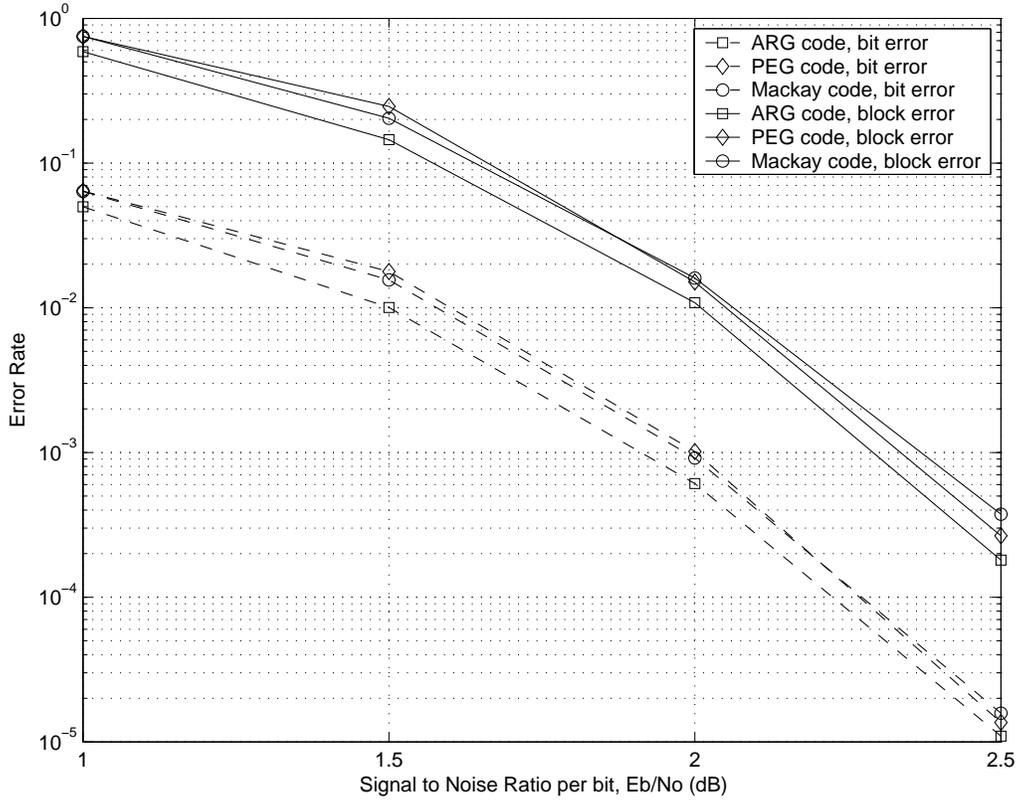}
	   \caption{Performance of ARG (1008,8,3) code}
       \label{fig1008}
   \end{center}
\end{figure*}

\begin{figure*}
  \begin{center}
	     \includegraphics[width=5.5in]{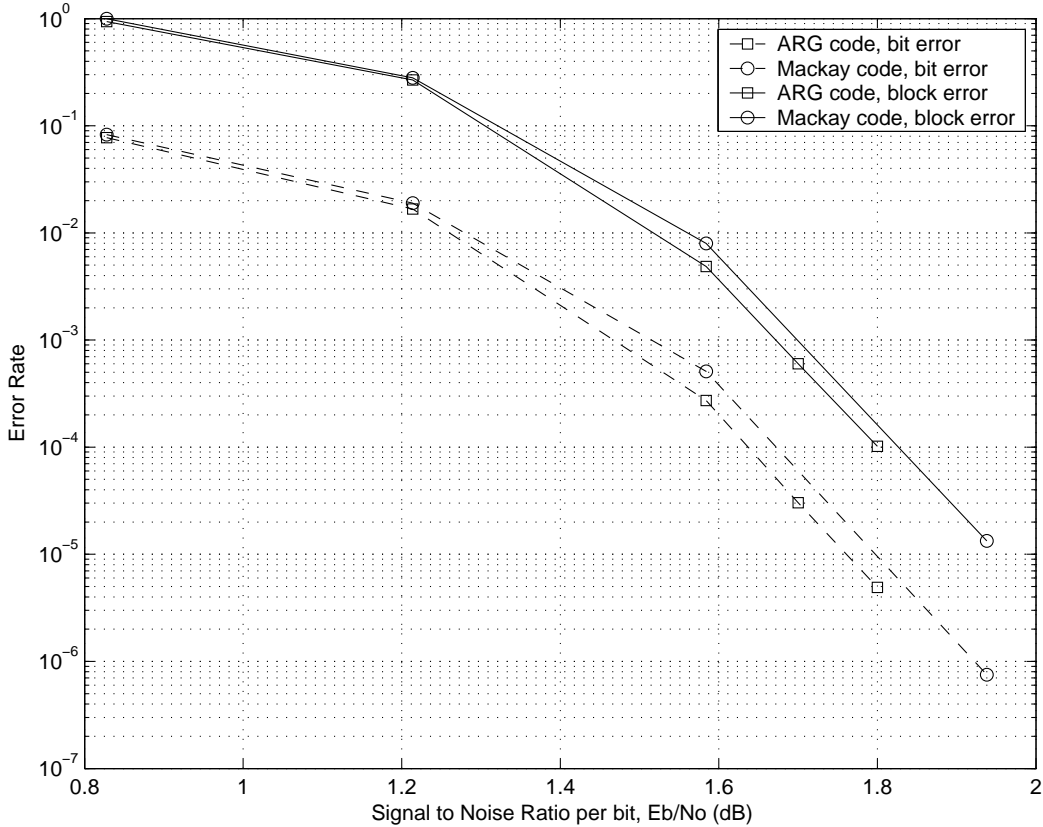}
	   \caption{Performance of ARG (4000,10,3) code}
       \label{fig1008}
   \end{center}
\end{figure*}

\begin{eqnarray}
m/2 \leq (pd+1)((pqd^{2})^{r}-1)/(pqd^2-1).
\end{eqnarray}
A similar argument for the case $x\in R$ and $y\in L$ for the 
addition of an even numbered edge yields the inequality:
\begin{eqnarray}
n/2 \leq (qd+1)((pqd^2)^{r}-1)/(pqd^2-1).
\end{eqnarray}
The statement of the theorem follows by noting that $m<n$
and taking the lower of the two bounds.

\end{proof}

The following table summarizes the minimum values of block length
required for achieving various values of girth and 
average left degree for codes of designed rate $1/2$. 
obtained by setting  $p=1$ and $q=2$.  
These values were obtained experimentally by varying the values of 
$d$ and $g$ given as input to the algorithm.
The minimum value of block length required
for achieving a given girth in actual experiments turns out 
to be lower than the bound proved in Theorem 2 indicating that 
the bound is not tight. \\


\begin{tabular}{|c| c| c|  c| c| c|}
\hline
\multicolumn{3}{|c|}{Code Parameters for rate 1/2 ARG codes} \\
\hline 
Left-degree     &Girth       &Block length	 \\   
(Average)       &            &               \\ \hline
3       &6       &40      \\  \hline 
4       &6       &80      \\  \hline
5       &6       &172     \\  \hline
3       &8       &252     \\  \hline
4       &8       &940     \\  \hline
3       &10      &1490    \\  \hline
\end{tabular}

\section{Complexity}
Assuming an adjacency list representation for the graph,
the selection of a farthest non-neighbour satisfying the degree
bound necessary during each edge addition may be performed by a
simple breadth first search in $O(n)$ time.  Since the total number
of edge additions is linear when $d$ is fixed constant, the overall 
construction complexity is $O(n^{2})$.

\section{Performance Simulations}

We shall refer to the code of block length $n$ defined by a 
Tanner graph of girth $g$ and average left degree $d$ as
an $(n,g,d)$ code.  Performance simulations for  
(504,8,3), (1008,8,3) and (4000,10,3) ARG codes of designed
rate $1/2$ (corresponding to $p=1$, $q=2$ in the algorithm)
are reported below.  The ARG codes perform slightly better than 
the MacKay and regular PEG codes of the same length \cite{Mc}.  
The bit and word error rate curves for the regular MacKay and PEG codes
of the same length are plotted together with those of the ARG 
code for easy reference.

\section{Conclusion}
We have presented an algorithm for constructing near-regular 
LDPC codes of large girth.  From a theoretical point of view,
the algorithm yields an asymptotic family with a provable
$\Omega(\log n)$ girth bound and quadratic complexity. 
The algorithm also gives good performance in practice in comparison with 
regular codes of the same length.  The problem of improving the girth 
bound remains open for further investigation.





\begin{thebibliography}{99}

\bibitem{Mc} D. J. C. MacKay.  Online Database of Low-Density
Parity-Check Codes.  Online: 
http://www.inference.phy.cam.ac.uk/mackay/CodesFiles.html.

\bibitem{Di}C. Di, D. Proietti, I. E. Telatar, T. J. Richardson and
R. Urbanke, "Finite length analysis of low-density parity-check
codes on the binary erasure channel," {\em IEEE Trans. Inf. Theory.},
Vol. 48, no. 6, pp. 1570-1579, June 2002.


\bibitem{Tian} T. Tian, C. Jones, J. D. Villasenor, R. D. Wesel,
"Construction of irregular LDPC codes with low error floors,"
{\em IEEE Intl. Conf. Comm.}, 2003, pp. 3125-3129.

\bibitem{Adi} A. Ramamoorthy, R. Wesel, "Construction of short block
length irregular LDPC codes," {\em  ICC 2004}, Paris, June 2004. 


\bibitem{Or} A. Orlitsky, R. Urbanke, K. Viswanathan, J. Shang,
"Stopping sets and girth of Tanner graphs," {\em ISIT 2002}, June 2002.


\bibitem{Tan} M. Tanner, "A recursive approach to low-complexity
codes," {\em IEEE Trans. Info. Theory}, Vol. 27, pp. 533-547, Sept 1981.


\bibitem{Gal} R. G. Gallager, "Low density parity-check codes," 
MIT Press, 1963.

\bibitem{Sunil} L. Sunil Chandran, "A High girth graph construction,"
{\em SIAM J. Discrete Math.}, Vol. 16, no. 3, pp. 366-370, 2003.


\bibitem{Sri1} R. M. Tanner, D. Sridhara, T. Fuja, 
"A class of group structured LDPC codes," {\em Proc. ICSTA 2001}, 
Ambleside, England, 2001.

\newpage

\bibitem{Sri2} R. M. Tanner, D. Sridhara, A. Sridharan, T. Fuja, 
D. J. Costello Jr., "LDPC block and convoluational codes based
on circulant matrices," {\em IEEE Trans. Info. Theory}, Vol. 50,
no.12, 2004.  .



\bibitem{Sri3} C. Kelley, D. Sridhara, "Pseudocodewords of Tanner Graphs," 
{\em arXiv: CS. IT/0504013}, April 2005.


\bibitem{Luby} M. G. Luby, M. Mitzenmacher, M. A. Shokrollahi,
D. Spielman, "Improved low density parity check codes using irregular
graphs and belief propagation," {\em IEEE Trans. Info. Theory},
Vol 47, pp.585-588, Feb. 2001.


\bibitem{Lin} Y. Kou, S. Lin, M. Fossorier, "Low density parity check
codes based on finite geometries: A rediscovery and new results," 
{\em IEEE Trans. Info. Theory}, Vol 47, pp.2711-2736, Nov. 2001.


\bibitem{Vas} B. Vasic, O. Milenkovic, "Combinatorial constructions 
of low density parity check codes for iterative decoding,"
{\em IEEE Trans. Info. Theory}, Vol 50, No. 6, June 2001.

\bibitem{Tang} H. Tang, J. Xu, Y. Mou, S. Lin, K. Abdel-Ghaffar,
"On algebraic construction of Gallager and circulant low-density
parity-check codes," {\em IEEE. Trans. Info. Theory}, Vol. 50,
No. 6, June 2004.


\bibitem{Sip} M. Sipser, D. A. Spielman, "Expander Codes,"
{\em IEEE. Trans. Info. Theory}, Vol. 42, pp.1710-1722, Nov 1996.

\bibitem{Ros} J. Rosenthal, P. O. Vontobel, "constructions of regular
and irregular LDPC codes using Ramanujan graphs and ideas from Margulis,"
{\em Proc. ISIT 2001}, p 4. June 2001.


\bibitem{Hu} Xiao-Yu Hu, "Regular and irregular progressive edge-growth
Tanner graphs," {\em IEEE. Trans. Info. Theory}, vol. 51, no. 1, 
Jan 2005 pp. 386-398.


\bibitem{Zem} G. Zemor, "On expander codes," 
{\em IEEE. Trans. Info. Theory}, vol. 47, no. 2, 
pp. 386-398,  Jan 2001. 

\end{thebibliography}
%

\end{document}